\documentclass[preprint,preprintnumbers,amsmath,amssymb]{revtex4}

\usepackage{graphicx}% Include figure files
\usepackage{dcolumn}% Align table columns on decimal point
\usepackage{bm}% bold math

\usepackage{amstext}
\usepackage[normalem]{ulem}

\usepackage{subfigure,axodraw}

\allowdisplaybreaks

\begin{document}

\title{The Eta decay into three neutral pions\\is mainly electromagnetic}% Force line breaks with \\

\author{A.  Nehme}
\email{anehme@ul.edu.lb}
\affiliation{%
Lebanese University\\
Faculty of Sciences at Nabatieh \\
Mount Amel, Lebanon
}%

\date{\today}% It is always \today, today,

\begin{abstract}
In the framework of Chiral Perturbation Theory including photons, we found
that the contribution of the photon exchange between two intermediate charged
Kaons to the slope parameter of the decay \(\eta\rightarrow 3\pi^{0}\) amounts
to \(-0.0221\pm 0.0034\). When compared with the experimental value, \(\alpha
=-0.0317\pm 0.0016\), on the one hand, and with the contribution of the up
and down quark mass difference, \(+0.013\pm 0.032\), on
the other hand, our result leads to the direct conclusion: \textit{The} \(\eta\rightarrow
3\pi^{0}\) \textit{decay \uline{cannot} be used to determine} \(m_{d}-m_{u}\).
\end{abstract}

\pacs{Valid PACS appear here}% PACS, the Physics and Astronomy
\maketitle

\section{Introduction}

The decay \(\eta\rightarrow 3\pi\), being forbidden by isospin symmetry,
takes contributions from the up and down quark mass difference, \(m_{d}-m_{u}\),
and from the electric charge, \(e\). Owing to Sutherland's theorem\cite{S66,BS68}, the latter contribution is suppressed with respect to the former. As a consequence,
the decay in question has been considered for a long while as the principal
source of information about \(m_{d}-m_{u}\). Theoretically, the decay was
studied in the framework of Chiral Perturbation Theory up to two loops\cite{GL85,BG07}.
The predicted value for the slope parameter in the neutral channel, \(\alpha_{\text{str}}=0.013\pm
0.032\) is in complete discrepancy with the observed value as quoted by the
Particle Data Group\cite{PDG10}, \(\alpha_\text{exp}=-0.0317\pm 0.0016\),
if we compare central values. The situation did not improve when taking into
account the electromagnetic interaction up to one loop\cite{BKW96,DKM09} and the suppression of the latter has been firmly confirmed. Recently, a
(partial) two-loop calculation\cite{NZ11} showed that the correction induced by the
diagram with a photon exchange between two intermediate charged pions on
the slope parameter amounts to \(\alpha_{\gamma}=0.0029\) and cannot be simply
neglected. Motivated by this promising result, we calculate in the present
work the correction induced by the exchange of a virtual photon between two
intermediate charged kaons on the slope parameter.

\section{The slope parameter}

\begin{figure}[ht]
\begin{center}
\subfigure[]{
\begin{picture}(100,60)(0,0)
\DashArrowLine(0,0)(20,30){3}
\ArrowLine(20,30)(0,60)
\ArrowLine(80,30)(100,0)
\ArrowLine(80,30)(100,60)
\GOval(50,30)(30,30)(0){1}
\Photon(50,60)(50,0){2}{6}
\end{picture}
} \quad 
\subfigure[]{
\begin{picture}(100,60)(0,0)
\DashArrowLine(0,0)(20,30){3}
\ArrowLine(20,30)(0,60)
\ArrowLine(80,30)(100,0)
\ArrowLine(80,30)(100,60)
\GOval(50,30)(30,30)(0){1}
\Photon(25,46)(76,46){1.5}{5}
\end{picture}
} \quad
\subfigure[]{
\begin{picture}(100,60)(0,0)
\DashArrowLine(0,0)(20,30){3}
\ArrowLine(20,30)(0,60)
\ArrowLine(80,30)(100,0)
\ArrowLine(80,30)(100,60)
\GOval(50,30)(30,30)(0){1}
\Photon(20,30)(80,30){2}{6}
\end{picture}
} \\
\subfigure[]{
\begin{picture}(100,60)(0,0)
\DashArrowLine(0,0)(20,30){3}
\ArrowLine(20,30)(0,60)
\ArrowLine(80,30)(100,0)
\ArrowLine(80,30)(100,60)
\GOval(50,30)(30,30)(0){1}
\PhotonArc(35,45)(21.2132,225,45){1.5}{8}
\end{picture}
} \quad
\subfigure[]{
\begin{picture}(100,60)(0,0)
\DashArrowLine(0,0)(20,30){3}
\ArrowLine(20,30)(0,60)
\ArrowLine(80,30)(100,0)
\ArrowLine(80,30)(100,60)
\GOval(50,30)(30,30)(0){1}
\PhotonArc(65,45)(21.2132,135,315){1.5}{8}
\end{picture}
} \\
\subfigure[]{
\begin{picture}(60,60)(0,0)
\DashArrowLine(0,0)(30,30){3}
\ArrowLine(30,30)(0,30)
\ArrowLine(30,30)(60,0)
\ArrowLine(30,30)(60,30)
\GOval(30,45)(15,15)(0){1}
\Photon(30,60)(30,30){1.5}{4}
\end{picture}
} \quad 
\subfigure[]{
\begin{picture}(60,60)(0,0)
\DashArrowLine(0,0)(30,30){3}
\ArrowLine(30,30)(0,30)
\ArrowLine(30,30)(60,0)
\ArrowLine(30,30)(60,30)
\GOval(30,45)(15,15)(0){1}
\Photon(15,45)(45,45){1.5}{4}
\end{picture}
} \quad
\subfigure[]{
\begin{picture}(100,60)(0,0)
\DashArrowLine(0,0)(20,30){3}
\ArrowLine(20,30)(0,60)
\ArrowLine(80,30)(100,0)
\ArrowLine(80,30)(100,60)
\Photon(20,30)(80,30){2}{6}
\GOval(20,30)(10,10)(0){0.9}
\GOval(80,30)(10,10)(0){0.9}
\end{picture}
}
\caption{\label{fig:diagrams}Feynman diagrams with photon exchange between
two intermediate charged Kaons. The dashed arrow represents the Eta, plain
arrows represent neutral pions, plain lines represent charged kaons, and
the wavy line represent a photon. The last diagram vanishes.}
\end{center}
\end{figure}
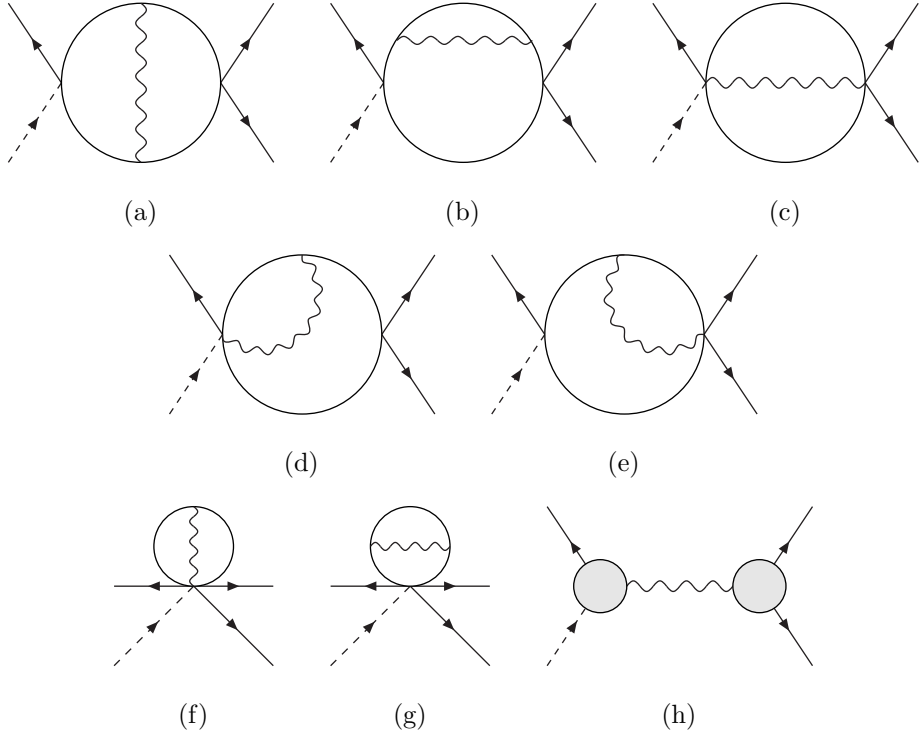

We follow the notation of \cite{NZ11} unless mentioned. The \(s\)-channel amplitude is written as
\begin{equation}\label{eq:amplitude}
M(s)=-\frac{\epsilon_{\eta\pi^{0}}}{3F_{\pi}^{2}}(M_{\eta}^{2}-M_{\pi^{0}}^{2})\left[ 1+\delta_{\text{str}}(s)+\delta_{\text{em}}(s)+\delta_{\gamma}(s) \right]
+\tilde \delta_{\text{em}}(s)+\tilde \delta_{\gamma}(s)\,,
\end{equation}
with \(\epsilon_{\eta\pi^{0}}\) is the \(\eta\)-\(\pi^{0}\) mixing angle
and reads
\begin{equation}
\epsilon_{\eta\pi^{0}}=\frac{\sqrt{3}}{4}\frac{m_{d}-m_{u}}{m_{s}-\hat m}+\text{higher-order
terms}\,.
\end{equation}
The strong interaction correction \(\delta_{\text{str}}\) contains both one-
and two-loop contributions\cite{GL85,BG07} and is of \(\mathcal{O} (p^{4})\) in the chiral counting. The one-loop electromagnetic corrections, \(\delta_{\text{em}}\)
and \(\tilde \delta _{\text{em}}\) are of respective chiral orders \(\mathcal{O} (e^{2})\) and \(\mathcal{O} (e^{2}p^{2})\), and have been calculated in \cite{DNT08,DKM09}
and \cite{BKW96,DNT08,DKM09}, respectively. The two-loop electromagnetic
correction \(\delta_{\gamma}\) is of \(\mathcal{O}(e^{2}p^{2})\) and has
been partially calculated in \cite{NZ11}. Finally, the two-loop electromagnetic
correction \(\tilde \delta_{\gamma}\) is of \(\mathcal{O}(e^{2}p^{4})\) and
is calculated in the present work. The corresponding Feynman diagrams with
a virtual photon exchange between two intermediate charged Kaons are sketched
in Fig.~\ref{fig:diagrams}.
The slope parameter is written as
\begin{equation}
\alpha =\alpha_{\text{str}}+\alpha_{\text{em}}+\tilde \alpha_{\text{em}}+\alpha_{\gamma}+\tilde \alpha_{\gamma}\,,
\end{equation}
where,
\begin{eqnarray}
\alpha_{\text{str}}
&=& \frac{1}{9}M_{\eta}^{2}(M_{\eta}-3M_{\pi^{0}})^{2}\text{Re}\;\delta_{\text{str}}''(s_{0})\,,
\\
\alpha_{\text{em}}
&=& \frac{1}{9}M_{\eta}^{2}(M_{\eta}-3M_{\pi^{0}})^{2}\text{Re}\;\delta_{\text{em}}''(s_{0})\,,
\\
\alpha_{\gamma}
&=& \frac{1}{9}M_{\eta}^{2}(M_{\eta}-3M_{\pi^{0}})^{2}\text{Re}\;\delta_{\gamma}''(s_{0})\,,
\\
\tilde \alpha_{\text{em}}
&=& -\frac{F_{\pi}^{2}}{3\epsilon_{\eta\pi^{0}}}\frac{M_{\eta}^{2}(M_{\eta}-3M_{\pi^{0}})^{2}}{M_{\eta}^{2}-M_{\pi^{0}}^{2}}\text{Re}\;\tilde\delta_{\text{em}}''(s_{0})\,,
\\
\tilde \alpha_{\gamma}
&=& -\frac{F_{\pi}^{2}}{3\epsilon_{\eta\pi^{0}}}\frac{M_{\eta}^{2}(M_{\eta}-3M_{\pi^{0}})^{2}}{M_{\eta}^{2}-M_{\pi^{0}}^{2}}\text{Re}\;\tilde\delta_{\gamma}''(s_{0})\,.
\end{eqnarray}
Diagrams of Fig.\ref{fig:diagrams} contribute to both \(\alpha_{\gamma}\)
and \(\tilde \alpha _{\gamma}\). Their contribution to the latter can be written, following \cite{NZ11} in terms of five Master Integrals as
\begin{eqnarray}
\tilde \alpha_{\gamma}
&=& -\frac{1}{1152\sqrt{3}}\frac{e^{2}}{\epsilon_{\eta\pi^{0}}F_{\pi}^{2}}\frac{M_{\eta}^{2}(M_{\eta}-3M_{\pi^{0}})^{2}}{M_{K}^{2}(M_{\eta}^{2}-M_{\pi^{0}}^{2})}\frac{1}{(4\pi
)^{D}(D-3)(D-4)}\times\nonumber\\
&& \frac{1}{s_{0}^{5}\sigma^{6}(s_{0})}\text{Re}\left(d_{1}J_{1}+d_{2}J_{2}+d_{3}JT+d_{4}J^{2}+d_{5}T^{2}\right)_{s=s_{0}}\,.
\end{eqnarray}
We found for the coefficients the following expressions,
\begin{eqnarray}
d_{1}
&=& (D-2) (512 (2 D (D (2 D-7)+3)+9) M_K^{10}\nonumber\\
&& -128 (D-2) (D (27 D-187)+294) s M_K^8\nonumber\\
&& -32 ((D (D (81 D-76)-1468)+2628) s^2\nonumber\\
&& +(2 D (D (2 D-7)+3)+9) (M_{\pi }^2+3 M_{\eta }^2){}^2) M_K^6\nonumber\\
&& +8 s (2 (D (D (217 D-2466)+8156)-8352) s^2\nonumber\\
&& -(D (D (21 D-167)+412)-312) (M_{\pi }^2+3 M_{\eta }^2){}^2) M_K^4\nonumber\\
&& +2 s^2 (12 (D-2) (D (79 D-426)+560) s^2\nonumber\\
&& -(D (D (27 D-272)+868)-888) (M_{\pi }^2+3 M_{\eta }^2){}^2) M_K^2\nonumber\\
&& +(D-2) s^3 (36 (D-2) D s^2-(D-6) (D-4) (M_{\pi }^2+3 M_{\eta }^2){}^2))\,,
\nonumber\\
d_{2}
&=& -6 (D-2)^2 (256 (D^2+D-9) M_K^8-64 (D (5 D-33)+48) s M_K^6\nonumber\\
&& -16 ((D (47 D-14)-468) s^2+(D^2+D-9) (M_{\pi }^2+3 M_{\eta }^2){}^2) M_K^4\nonumber\\
&& +4 s (12 (D-4) (20 D-43) s^2-(D (7 D-51)+84) (M_{\pi }^2+3 M_{\eta }^2){}^2) M_K^2\nonumber\\
&& +36 (D-2) D s^4-(D-6) (D-4) s^2 (M_{\pi }^2+3 M_{\eta }^2){}^2)\,, \nonumber\\
d_{3}
&=& -256 (D (D (7 D-46)+115)-108) M_K^{10}\nonumber\\
&& -256 (D-2) (D ((D-21) D+91)-108) s M_K^8\nonumber\\
&& +16 ((D (D (D (D (D+92)-751)+1742)-1028)-336) s^2\nonumber\\
&& +(D (D (7 D-46)+115)-108) (M_{\pi }^2+3 M_{\eta }^2){}^2) M_K^6\nonumber\\
&& -16 s ((D (D (D (D (3 D+107)-1195)+3646)-3236)-528) s^2\nonumber\\
&& -(D (D ((D-14) D+67)-149)+132) (M_{\pi }^2+3 M_{\eta }^2){}^2) M_K^4\nonumber\\
&& +s^2 (12 (D-2) (D (D (D (3 D+14)-397)+1456)-1520) s^2\nonumber\\
&& -(D (D (D ((D-4) D-67)+566)-1592)+1536) (M_{\pi }^2+3 M_{\eta }^2){}^2) M_K^2\nonumber\\
&& +(D-2)^2 s^3 (36 (D-2) D s^2-(D-6) (D-4) (M_{\pi }^2+3 M_{\eta }^2){}^2)\,,
\nonumber\\
d_{4}
&=& -2 M_K^2 (36 (D-3)^2 (D-2) (3 D-8) s^5-(D-3) (2048 (D-2) M_K^{10}\nonumber\\
&& +512 (D-6) (D-2) s M_K^8+64 (D-2) ((D-6) (D-5) s^2-2 (M_{\pi }^2+3 M_{\eta }^2){}^2) M_K^6\nonumber\\
&& -16 s ((D (D (15 D-23)-108)+128) s^2+2 (D-6) (D-2) (M_{\pi }^2+3 M_{\eta }^2){}^2) M_K^4\nonumber\\
&& +4 s^2 (24 (D-3) (D (3 D-1)-20) s^2-(D-6) (D-5) (D-2) (M_{\pi }^2+3 M_{\eta }^2){}^2) M_K^2\nonumber\\
&& +(D-5) (D-4) (3 D-8) s^3 (M_{\pi }^2+3 M_{\eta }^2){}^2))\,, \nonumber\\
d_{5}
&=& -(D-2) (384 (3 D ((D-4) D+1)+20) M_K^8\nonumber\\
&& +32 (D (D ((D-48) D+327)-776)+616) s M_K^6\nonumber\\
&& -8 (2 (D (D (6 D^2+75 D-430)-322)+1976) s^2\nonumber\\
&& +3 (3 D ((D-4) D+1)+20) (M_{\pi }^2+3 M_{\eta }^2){}^2) M_K^4\nonumber\\
&& +2 s (36 (D (D (D (D+18)-217)+644)-584) s^2\nonumber\\
&& -(D (D (D (D+12)-201)+724)-776) (M_{\pi }^2+3 M_{\eta }^2){}^2) M_K^2\nonumber\\
&& +108 (D-2)^2 D s^4-(D-4)^2 (3 D-8) s^2 (M_{\pi }^2+3 M_{\eta }^2){}^2)\,.
\end{eqnarray}
The analytic expressions for the integrals can be found in \cite{NZ11} with
\(M_{\pi}\) replaced by \(M_{K}\). Concerning the \(J\) integral, it should
be replaced by the following expression,
\begin{equation}
J=i\left\{ 1-M_{\pi}^{D-4}\Gamma (1-D/2)-2\left( \frac{4M_{K}^{2}}{s}-1 \right)^{1/2}\arctan\left( \frac{4M_{K}^{2}}{s}-1 \right)^{-1/2}\right\}\,.
\end{equation}

\section{Results and conclusions}

We first expand \(\tilde\alpha_{\gamma}\) around \(D=4\) and then use the following
numerical values
\begin{equation}
e^{2}=\frac{4\pi}{137.04}\,, \quad (F_{\pi},M_{\pi^{0}},M_{K},M_{\eta})=(92.42,\,
139.57,\, 493.68,\, 547.30)\;\text{MeV}\,.
\end{equation}
For the mixing angle, we use the value\cite{EMNP00},
\begin{equation}
\epsilon_{\eta\pi^{0}}=0.013\pm 0.032\,.
\end{equation}
We find that
\begin{equation}\label{eq:result}
\tilde \alpha _{\gamma}=-0.0221\pm 0.0034\,.
\end{equation}
On the other hand, we found that the contribution of the diagrams in Fig.\ref{fig:diagrams}
to \(\alpha_{\gamma}\) is equal to \(-0.0005\). This, together with the pion contribution calculated in \cite{NZ11}, give a total of
\begin{equation}
\alpha_{\gamma}=0.0024\,.
\end{equation}
Adding now the whole electromagnetic contribution to the slope, we get
\begin{equation}
\alpha_{e^{2}}=\alpha_{\text{em}}+\alpha_{\gamma}+\tilde \alpha_{\text{em}}+\tilde \alpha_{\gamma}=-0.0203\pm 0.0034\,.
\end{equation}
Note that the uncertainty comes from the mixing angle only. The obtained
result contradicts all the ``classical wisdom'' based on the suppression
of the electromagnetic contribution to the slope parameter. Finally, the
main result of the present work, Eq.~(\ref{eq:result}), concerns only Kaon loops with one virtual photon exchanged. In order to obtain finite and scale
independent amplitude, one needs to perform three additional calculations:
\begin{enumerate}
\item 
Two-loop diagrams without photons, but with mass difference between charged
and neutral mesons (pions and kaons);
\item
One-loop diagrams with counterterms of \(\mathcal{O}(e^{2}p^{2})\) in the
vertices;
\item
Tree-level diagram with \(\mathcal{O} (e^{2}p^{4})\) counterterms.
\end{enumerate}
In order to make the last calculation, the construction of a chiral Lagrangian
is necessary.


\begin{thebibliography}{99}

\bibitem{S66}
D. G. Sutherland, Phys. Lett. \textbf{23}, 384 (1966).

\bibitem{BS68}
J. S. Bell and D. G. Sutherland, Nucl. Phys. B \textbf{4}, 315 (1968).

\bibitem{GL85}
J. Gasser and H. Leutwyler, Nucl. Phys. \textbf{B250}, 539 (1985).

\bibitem{BG07}
J. Bijnens and K. Ghorbani, JHEP \textbf{11}, 030 (2007).

\bibitem{PDG10}
K. Nakamura \textit{et al.}, JP G \textbf{37}, 075021 (2010).

\bibitem{BKW96}
R. Baur, J. Kambor, and D. Wyler, Nucl. Phys. \textbf{B460}, 127 (1996).

\bibitem{DKM09}
C. Ditsche, B. Kubis, and U.-G. Meissner, Eur. Phys. J. C \textbf{60}, 83
(2009).

\bibitem{NZ11}
A. Nehme and S. Zein, hep-ph/1106.0915.

\bibitem{DNT08}
A. Deandrea, A. Nehme, and P. Talavera, Phys. Rev. D \textbf{78}, 034032
(2008).

\bibitem{EMNP00}
G. Ecker, G. Muller, H. Neufeld and A. Pich, Phys. Lett. B \textbf{477},
88 (2000).

\end{thebibliography}
\end{document}